\documentclass[twocolumn,prl,amsmath,amssymb,floatfix]{revtex4}
\usepackage{graphicx}

\begin{document}

\title{d-dimensional oscillating scalar field lumps and the dimensionality of space}

\author{Marcelo Gleiser}
\email{gleiser@dartmouth.edu}

\affiliation{Department of Physics and Astronomy, Dartmouth College,
Hanover, NH 03755, USA}

\date{\today}

\begin{abstract}
Extremely long-lived, time-dependent, spatially-bound scalar field 
configurations are shown to exist in $d$ spatial dimensions
for a wide class of polynomial interactions parameterized as
$V(\phi) = \sum_{n=1}^h\frac{g_n}{n!}\phi^n$. Assuming
spherical symmetry and if $V''<0$ for a range of values of $\phi(t,r)$, 
such configurations exist if: i) spatial dimensionality is
below an upper-critical dimension $d_c$; ii) their radii are above 
a certain
value $R_{\rm min}$. Both $d_c$ and $R_{\rm min}$ are uniquely determined by 
$V(\phi)$.
For example, symmetric double-well potentials only sustain such 
configurations if $d\leq 6$ and 
$R^2\geq d[3(2^{3/2}/3)^d-2]^{-1/2}$. 
Asymmetries may modify the value of $d_c$.
All main analytical results are confirmed numerically. Such objects 
may offer novel ways to probe the dimensionality of space.

\end{abstract}

\maketitle

\section{Introduction}
It is well known that many physical systems may be efficiently modeled in a 
reduced number of spatial dimensions. In particular, certain static solutions of
nonlinear classical field equations exhibiting solitonic behavior
have been used to describe a wide variety of phenomena, ranging from hydrodynamics
and condensed matter physics \cite{solitons} to  
relativistic field theories \cite{rajaraman}. At the opposite
extreme, the possibility that the four fundamental interactions may be unified
in theories with extra spatial dimensions has triggered research
on the existence of static nonperturbative solutions of nonlinear field theories in
more than three spatial dimensions \cite{callan} \cite{larsen-khan}. 
These extra dimensions may be compact
and much smaller than the usual three dimensions of space, as in Kaluza-Klein (KK)
theories \cite{kaluza-klein}, or they may
be infinitely large, as in the recent Randall-Sundrum (RS)
proposal, where gravity (and possibly other fields)
can leak into the extra dimension transverse
to the 3-dimensional brane where matter and gauge fields propagate
\cite{randall-sundrum}.
There have been many variants of the RS proposal \cite{rubakov}, 
including some with more than 
one brane \cite{gregory} or with more than a single
large extra dimension \cite{arkani-hamed}. In either the KK or the RS scenarios,
there is plenty of motivation to study $d$-dimensional nonperturbative field
configurations with a large number of quanta $N$, 
even if their direct production in particle colliders is probably 
exponentially suppressed as $\sim \exp[-N]$. [There has been much interest recently
in the possibility that extra dimensions could produce signatures observable in 
collider experiments
\cite{antoniadis} \cite{giudice} \cite{quiros} \cite{allanach} \cite{muck}
\cite{hossenfelder}. Although I will not examine if
the objects of the present work could be produced in future collider experiments,
the possibility should be kept in mind, especially with more realistic models
involving couplings between several fields.] 

In this work, I argue that long-lived
{\it time-dependent} $d$-dimensional scalar field lumps  -- {\it oscillons} --
can exist in a wide class of
models, much wider than their static (solitonic) counterparts.
Furthermore, I show that they exist only below a certain critical number
of spatial dimensions,
which is uniquely determined by the field's self-interactions. 
If the fundamental gravity scale is $M$, the associated
length scale of the extra dimensions is $R_{\rm KK}
\simeq (M_{\rm Pl}/M)^{2/(d-3)}M^{-1}$. 
[$d-3 \geq 1$ is the number of extra dimensions.] Thus, if $M\sim 1$ TeV,
$R_{\rm KK}\sim 10^{32/(d-3)} 10^{-17}$ cm. For $d\geq 5$, this scenario is still
acceptable by current tests of Newton's gravitational law \cite{rubakov}. 
Even though oscillons ultimately decay, their lifetimes are long enough
to produce significant effects: their demise occurs in very short time-scales
and hence would appear, in the scenario with large but compact
extra dimensions, as a sudden burst of particles from a small region.
If these particles are quanta of the scalar field, their masses
would satisfy, assuming
maximally-symmetric internal dimensions,
$p_{\mu}p^{\mu} + n^2/R^2_{\rm KK} = V''(\phi_v)^{1/2}$, where
$\mu = 0,1,2,3$, and
$V''(\phi_v)^{1/2}$ is the tree-level 
mass of vacuum excitations satisfying the $d$-dimensional Klein-Gordon
equation. The key point here is that
since the mass and size 
of $d$-dimensional oscillons are uniquely
determined by the number of spatial 
dimensions and their interaction potential, they can serve as probes 
to the dimensionality of space. A $d$-dimensional
oscillon hypothetically appearing at the TeV energy scale will have a
typical size of order $d$ TeV$^{-1}\sim 10^{-17}$ cm, always much smaller than 
$R_{\rm KK}$. 

So far, most work in either a reduced or increased number of spatial dimensions
has focused on {\it static} solutions involving real scalar fields or
scalar fields coupled to other fields. A recent example is the work by Bazeia et al.,
where $d$-dimensional spherically-symmetric topological defects were found for models 
with potential $U(x^2;\phi)= f(x^2)V(\phi)$ \cite{bazeia}. The particular
choice of potential is needed to evade Derrick's theorem, which forbids the existence of
non-trivial static solutions for real scalar fields in more than one 
spatial dimension \cite{derrick}.
When time-dependence is introduced, it is often as
a general phase of a complex scalar field, $\phi(x,t) =
\varphi(x)\exp[-i\omega t]$, such that the equations still allow for localized
solutions with static spatial profiles. Nontopological solitons \cite{nts} and
$Q$-balls \cite{qballs} are well-known examples of such configurations. There are
exceptions, though. Breathers
in one dimension \cite{breathers}, and oscillons in two \cite{gleiser-sornborger}
\cite{gleiser-howell} \cite{adib} and three \cite{gleiser} \cite{copeland} \cite{honda} 
\cite{hsu} \cite{riotto}
are spatially-bound, time-dependent scalar field configurations which are remarkably
long-lived. They are found in many physical systems and models, including vibrating
grains, Josephson junctions, nonlinear Schr\"odinger equations, Ginzburg-Landau models, 
and certain relativistic $\phi^4$ models, to name a few examples. As will be seen, 
they also exist in higher-dimensional models for a wide class of polynomial
interactions. 

\section{Scalar field dynamics in d-dimensions}

The line element for flat $d+1$-dimensional spacetime is 
$ds^2=\eta_{MN}dx^{M}dx^{N}$,
where $M,N=0,1,2,...,d$ and $\eta_{MN} = {\rm diag}(+,-,-....,-)$. 
I'm only concerned here with objects which may exist in the full
$d$ dimensions. 
Thus, their typical size $R_{\rm min}$ will have to
satisfy $R_{\rm min} \ll R_{\rm KK}$, where $R_{\rm KK}$ is the linear size of the extra
dimensions.

Since any deformation away from spherical symmetry leads to more energetic 
configurations \cite{adib} \cite{qballs}, I will consider only
spherically-symmetric configurations, $\phi(t,r)$. As such, 
the $d$-dimensional spatial volume 
element can be written as $d^dx = c_dr^{(d-1)}dr$, where $c_d=2\pi^{d/2}/\Gamma(d/2)$ 
is the surface area of a $d$-dimensional sphere of unit radius. 
The Lagrangian can be written as
\begin{equation}
\label{lagrangian1}
L = c_d\int r^{(d-1)}dr \left( \frac{1}{2}\dot\phi^2 - 
\frac{1}{2}\left (\frac{\partial\phi}{\partial r}\right )^2
-V(\phi)\right)~,
\end{equation}
\noindent
where a dot means time derivative.

Previous results in $d=2$ and $d=3$ have shown that oscillons are well-approximated by 
configurations with a general Gaussian profile \cite{copeland} \cite{gleiser-sornborger}
\cite{gleiser-howell}.
In fact, oscillons can be viewed as localized configurations modulated by nonlinear
oscillations on the field's amplitude. I will thus treat the amplitude as a function of 
time, keeping the radius $R$ constant. 
The fact that the analytical results to be obtained are
verified numerically to great accuracy 
confirms that this approximation is adequate for this work's purpose.
Oscillons are thus modeled as
\begin{equation}
\label{ansatz}
\phi(t,r) = \left [\phi_c(t) - \phi_v\right ]\exp[-r^2/R^2] + \phi_v~,
\end{equation}
\noindent
where $\phi_c(t)$ is the core value of the field [$\phi(t,r=0)$],
and $\phi_v$ is its asymptotic value at spatial infinity, 
determined by $V(\phi)$. Thus, one condition on $V(\phi)$ is that
$\partial^2V/\partial\phi^2|_{\phi_v} > 0$. Note also that the equation of motion for
$\phi(t,r)$ imposes that $\phi'(r=0) = 0$, a condition satisfied by the 
Gaussian ansatz above. 

Since many applications involve polynomial potentials, we will write $V(\phi)$ as
\begin{equation}
\label{potential}
V(\phi) = \sum_{j=1}^{h} \frac{g_j}{j!}\phi^j - V(\phi_v)~,
\end{equation}
\noindent
where the $g_j$'s are constants. The vacuum energy $V(\phi_v)$ is
subtracted from the potential to avoid spurious divergences upon 
spatial integration.

Substituting the ansatz of eq. (\ref{ansatz}) into eq. (\ref{lagrangian1}), 
one may perform the spatial integrations to obtain,
\begin{eqnarray}
\label{lagrangian2}
L[A,R,\dot A] &=& \left(\frac{\pi}{2}\right)^{d/2}R^d \left[ \frac{{\dot A}^2}{2}
- \frac{d}{2R^2}A^2 \right. \nonumber \\
&-& \left. \sum_{n=2}^{h} \left(\frac{2}{n}\right)^{d/2}\frac{1}{n!}
V^n(\phi_v)A^n \right]~, 
\end{eqnarray}
\noindent
where $V^n(\phi_v)\equiv \frac{\partial^n  V(\phi_v)}{\partial \phi^n}$,
and I introduced the amplitude $A(t)\equiv \phi_c(t)-\phi_v$. 
Note that the sum in the last term starts at $n=2$. This is
due to the fact that, by definition, $\frac{\partial  V(\phi_v)}{\partial \phi}=0$.

\section{Upper Critical Dimension for Oscillons}

From the Lagrangian in
eq. (\ref{lagrangian2}) one obtains the equation of motion for $A(t)$:
\begin{equation}
\label{Rconst}
\ddot A = -\frac{d}{R^2}A - \sum_{n=2}^h \left(\frac{2}{n}\right)^{d/2}
\frac{1}{(n-1)!}V^n(\phi_v)A^{n-1}~.
\end{equation}
If $V(\phi)=0$, the amplitude undergoes harmonic
oscillations with constant frequency $\omega^2 = \frac{d}{R^2}$. This behavior is due to
the surface term that resists any displacement from equilibrium, $A=0$. Note 
that since
the Lagrangian was integrated over all space, the model cannot
describe the fact that the configuration decays 
by radiating its energy 
to spatial infinity. If needed, one could include a phenomenological term $\gamma\dot A$
in order to mimic this effect (such as $\gamma \sim t^{-3/2}$ in $d=3$ \cite{copeland}), 
although this is not relevant for the present work.

To examine the stability of the motion, I expand the amplitude as
$A(t) = A_0(t)+\delta A(t)$. Linearizing eq. (\ref{Rconst}),
\begin{eqnarray}
\label{linear}
\delta\ddot A &=& -\left[\frac{d}{R^2}+\sum_{n=2}^h \left(\frac{2}{n}\right)^{d/2}
\frac{1}{(n-2)!}V^n(\phi_v)A_0^{n-2}\right]\delta A \nonumber \\
&\equiv& -\omega^2(R,A_0)\delta A~,
\end{eqnarray}
\noindent
where I introduced the effective frequency $\omega^2$ in the last line. 
Instabilities occur if $\omega^2 < 0$. Long-lived oscillons are only possible
if the oscillations above the vacuum with amplitude $A(t)$ probe
regions of $V''<0$ for a sustained period of time \cite{gleiser} \cite{copeland}. 
This requires $\omega^2 <0$ for oscillons to exist.

I proceed by deriving several results from the expression for $\omega^2$.
First, it is useful to write it fully as,
\begin{eqnarray}
\label{omega2}
&&\omega^2(R,A_0) = \frac{d}{R^2} + V''(\phi_v) + 
\left(\frac{2}{3}\right)^{d/2}V'''(\phi_v)A_0 \nonumber \\
&+& \left(\frac{1}{2}\right)^{d/2}V^{IV}(\phi_v)A_0^2 +
\left(\frac{2}{5}\right)^{d/2}\frac{1}{3!}V^{V}(\phi_v)A_0^3+ \cdots
\end{eqnarray}
\noindent
Note that since $A_0$ is a function of time, eq. (\ref{linear})
is in the form of a Mathieu equation. Although the time dependence is
crucial in the study of oscillon dynamics (c.f. ref. \cite{gleiser-howell}),
it will not be relevant here.

\subsection{Quadratic potentials}

If $V(\phi)$ is quadratic, only the first two terms on the RHS of
eq. (\ref{omega2}) contribute to $\omega^2$.
For $V''(\phi_v) > 0$, $\omega^2 >0$ and no instability occurs \cite{copeland}. 
The field will simply
undergo damped oscillations about $A=0$. If $V''(\phi_v)<0$, instabilities are possible
only for $d/R^2 < |V''(\phi_v)|$ or
\begin{equation}
\label{instabquad}
R \geq \left( \frac{d}{|V''(\phi_v)|} \right)^{1/2}~.
\end{equation}
This is the well-known spinodal instability bound \cite{goldenfeld},
where the critical wavelength $\lambda_c$ is related to the initial size of the configuration,
$R$, by $R = (\sqrt{d}/2\pi)\lambda_c$. When the condition for instability is satisfied,
the amplitude $A(t)$ will grow without bound as the field rolls down the potential. This
behavior will depend on the damping term, that is, on the rate at which the lump radiates
energy to spatial infinity. If the condition eq. (\ref{instabquad}) is {\it not} satisfied,
the amplitude will describe damped oscillations about the origin. Note that this
can only happen in a field theory, since the gradient term is needed to allow for
$\omega^2 >0$ even if $V''(\phi_v) <0$. In effect, the gradient term stabilizes
what would have been an unstable configuration. This mechanism is favored as $d$ 
increases, as one would expect.

Gaussian-shaped bubbles with quadratic potentials are thus
short-lived, not a surprising result \cite{copeland}: 
oscillons owe their longevity to nonlinearities in the potential.
Furthermore, as it is proven next, a necessary condition for their
existence is that the potential satisfies
$V''(A) < 0$ for at least a range of amplitudes. This is
also true for solitons, which do not exist for potentials with positive concavity,
e.g., $V(\phi) = \phi^2 + \phi^4$. This necessary condition, however, is not sufficient to
guarantee the existence of oscillons.

\subsection{Cubic potentials}

If $V(\phi)$ is cubic, the first thing to notice is that since parity is broken, $V(\phi)$
will always have an inflection point at $\phi_{\rm inf} = -g_2/g_3.$ 
The choice of vacuum will
depend on the sign of $g_2$: for $g_2>0$, $\phi_v=0$; for $g_2 <0$, $\phi_v = -2g_2/g_3$.
In either case, the condition $\omega^2 <0$ will be satisfied whenever $A_0$ has opposite 
sign to $g_3$ {\it and} for values of $R\geq R_{\rm min}$ as summarized in
Table 1. 

{\large
\begin{center}
\begin{tabular}{||c||c|c||}\hline
 $~$ & $g_3>0$ & $g_3<0$ \\ \hline
 $g_2>0$ & $\frac{d}{g_3\left[\left(\frac{2}{3}\right)^{d/2}|A_0| + \phi_{\rm inf}\right]}$ & 
 $\frac{d}{|g_3|\left[\left(\frac{2}{3}\right)^{d/2}A_0 - \phi_{\rm inf}\right]}$ \\ \hline
 $g_2<0$ & $\frac{d}{g_3\left[\left(\frac{2}{3}\right)^{d/2}|A_0| - \phi_{\rm inf}\right]}$ & 
 $\frac{d}{|g_3|\left[\left(\frac{2}{3}\right)^{d/2}A_0 + \phi_{\rm inf}\right]}$ \\ \hline
\end{tabular}
\end{center} 
}
\noindent
Table 1: Values of $R^2_{\rm min}$ for different couplings of the cubic potential model.

Since $R^2>0$, the amplitudes must satisfy, for any of the cases in Table 1,
\begin{equation}
\label{cond-cubic}
|A_0| \geq \left(\frac{3}{2}\right)^{d/2}|\phi_{\rm inf}|~,
\end{equation}
\noindent
showing that only fluctuations probing deep into the spinodal region of the potential will be
able to sustain long-lived oscillons. It is also clear that the higher the dimensionality
the larger the amplitudes need to be. However, for cubic potentials, 
as long as the conditions
above are satisfied, long-lived oscillating lumps can exist in any number of dimensions. 
This result does not
hold for arbitrary polynomial potentials, as we see next.

\subsection{Quartic potentials}

For quartic potentials, the condition for the existence of oscillating lumps becomes,
\begin{eqnarray}
\label{quartic-omega}
\omega^2(R,A_0) &\leq& \frac{d}{R^2}+V''(\phi_v)+
\left(\frac{2}{3}\right)^{d/2}V'''(\phi_v)A_0 \nonumber \\
&+&\frac{1}{2}\left(\frac{1}{2}\right)^{d/2}V^{IV}(\phi_v)A_0^2~.
\end{eqnarray}
Results are sensitive to the sign of $V^{IV}(\phi_v) = g_4$. Let me first examine the case
for $g_4>0$: $\omega^2$ is a parabola with positive concavity. 
Thus, if $\omega^2<0$ at its minimum,
the condition is satisfied for a range of amplitudes. The minimum is at 
${\bar A}_0 = -(4/3)^{d/2}V'''/V^{IV}$, and
\begin{equation}
\label{omega-min}
\omega^2(R,{\bar A}_0) = \frac{d}{R^2}+V''
-\left(\frac{2}{3}\right)^d2^{(d-2)/2}\frac{(V''')^2}{V^{IV}}~.
\end{equation}
{}For $\omega^2<0$,
\begin{equation}
\label{minimum-radius}
R^2 \geq \frac{d}{\left[\frac{1}{2}\left(\frac{2^{3/2}}{3}\right)^d
\frac{(V''')^2}{V^{IV}}-V''\right]}~.
\end{equation}
So, as in the case for cubic potentials,
oscillating lumps can only exist for radii above a critical size. 
This has been observed numerically
for double-well potentials in $d=2$ \cite{gleiser-sornborger} and 
$d=3$ \cite{gleiser}. 
Notice also that 
since the denominator must be positive definite, this condition
imposes both a restriction on the potential {\it and} an upper 
critical dimension for oscillons:
\begin{equation}
\label{max-d}
d\leq {\rm Int}\left[\frac{ \ln 2\frac{V''V^{IV}}{(V''')^2}}
{\ln \left(\frac{2^{3/2}}{3}\right)}\right]~,
\end{equation}
\noindent
where ${\rm Int}$ means the integral part of. Equality defines the upper critical
dimension for the existence of oscillons, $d_c$.
Since $\frac{2^{3/2}}{3}<1$, the potential must satisfy
\begin{equation}
\label{cond-pot}
2\frac{V''V^{IV}}{(V''')^2} < 1~,
\end{equation}
\noindent
for the bifurcation instability on the radius to occur and, 
thus, for oscillating lumps to exist.

If $g_4<0$, $\phi_v=0$ for stability. In
this case, $g_2>0$ (the sign of $g_3$ is irrelevant) 
and there will always be an amplitude $|A_0|$ 
large enough so that $\omega^2 <0$,
\begin{eqnarray}
A_0 &=& 2^{d/2}\left(\frac{2}{3}\right)^{d/2}\frac{g_3}{|g_4|}\times \\ \nonumber
&& \left[1\pm \left(1+3^d2^{-3d/2}2\left(g_2+\frac{d}{R^2}\right)\frac{|g_4|}{g_3^2}\right)\right]~.
\end{eqnarray}
Comparing this amplitude with the value for the inflection point, $\phi_{\rm inf} = 
g_3/|g_4|\left[1\pm \left(1+2g_2|g_4|/g_3^2\right)\right]$, it's easy to see that
$|A_0|>|\phi_{\rm inf}|$, that is, oscillons only exist if the amplitudes go beyond
the inflection point.

\section{Application: Symmetric and asymmetric double-well potentials in d-dimensions}

{}For a symmetric double-well (SDW) potential with $g_4>0$, $V(\phi) = 
\frac{\lambda}{4}\left(\phi^2-\phi_v^2\right)^2$, the coefficients of the general expression
eq. (\ref{potential}) are;
$h=4,~~g_1=g_3=0,~~g_2=-\lambda\phi_v^2,~~g_4=6\lambda$,
and thus the various derivatives at $\phi_v$ are 
$V'(\phi_v)=0,~~V''(\phi_v)=2\lambda\phi_v^2,~~
V'''(\phi_v)=6\lambda\phi_v,~~V^{IV}(\phi_v)=6\lambda$.
The necessary condition eq. (\ref{cond-pot}) is satisfied,
$2V''V^{IV}/(V''')^2 = 2/3 <1$. From eq. (\ref{minimum-radius}), oscillons will exist
if the radius is larger than [the radius can be made dimensionless with the
scaling $R = R'/\sqrt{\lambda}\phi_v$ for any $d$]
\begin{equation}
\label{minrad SDW}
R^2 \geq  \frac{d}{\left[3\left(\frac{2^{3/2}}{3}\right)^d - 2\right]}~.
\end{equation}
\noindent
For $d=2$, $R_{\rm min} = \sqrt{3}$ \cite{gleiser-sornborger}. For
$d=3$, $R_{\rm min} \simeq 2.42$ \cite{copeland}. 
The expression predicts that, e.g., for $d=6$, $R_{\rm min} \simeq 7.5$. 
Note also that the SDW has an upper critical dimension of $d_c=6$: 
from eq. (\ref{max-d}), $d\leq {\rm Int}\left[\ln (2/3)/\ln (2^{3/2}/3)
\right] = 6$.

I have confirmed these results numerically, using a
leap-frog method fourth-order accurate in space. The
lattice spacing was $\delta r = 0.01$ and the time step
$\delta t = 0.001$. Energy was conserved to better than
one part in $10^5$. The program solves the $d$-dimensional
Klein-Gordon equation in spherical coordinates with
initial condition set to be a Gaussian bubble with radius $R$
and $\phi_c = 1$ and $\phi_v=-1$. [One can vary the initial
profile and parameters at will; if the conditions for the appearance of
oscillons are satisfied, the field will evolve into
an oscillon configuration, since
it is an attractor in field-configuration space \cite{adib} \cite{copeland}. 
Furthermore, oscillons have
been shown to emerge even from thermal initial states \cite{gleiser-howell}.]
The program reproduced results from refs. \cite{gleiser-sornborger}
and \cite{gleiser} in $d=2$ and $d=3$, respectively. In Fig. 1,
I show the energy within a shell of radius $R_{\rm shell} = 10R$
as a function of time for $d=4$ ($R_{\rm min} \simeq 3.29$). 
The approximately flat plateaus denote
oscillons in 5-dimensional spacetime. As in $d=2$ and $d=3$, there is a
range of values of $R$ that lead to oscillons: larger values produce
configurations that decay without settling into an oscillon.
\begin{figure}[t]
\includegraphics[width=3in,height=3in]{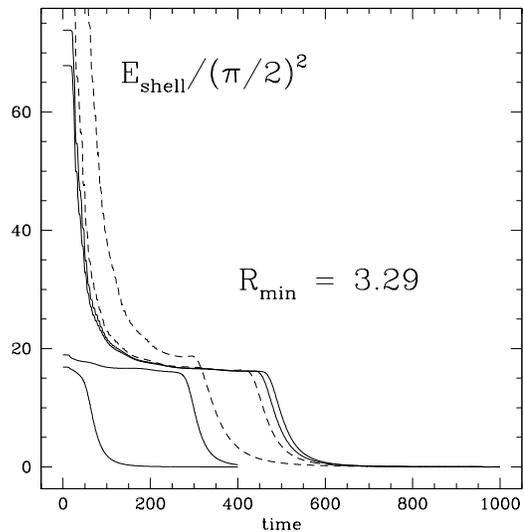}
\caption[oscillons for d=4]
{Time evolution of the energy within a shell of radius $R_{\rm shell} =10R$
in $d=4$. From left to right: continuous lines are for $R_{\rm eff} = 3.17, 3.29, 4.81,
4.93$; dashed lines are for  $R_{\rm eff} = 5.87, 5.17$. The plateaus denote oscillons.}
\label{d4_oscil}
\end{figure}

The effective radius of the configuration,
the one checked against the predictions of eq. \ref{minrad SDW},
is computed as the normalized
first moment of the energy distribution,
\begin{equation}
\label{Reff}
R_{\rm eff} = \frac { \int r^d dr \left[\frac{1}{2}\dot \phi^2
+ \frac{1}{2}\phi'2 +V(\phi)\right] }{\int r^{(d-1)} dr
\left[\frac{1}{2}\dot \phi^2
+ \frac{1}{2}\phi'2 +V(\phi)\right] }~.
\end{equation}
In Fig. 2,
I show similar results for $d=6$, again confirming the prediction of
eq. (\ref{minrad SDW}). [The very narrow low-energy plateaus seen in Fig. 2
seem to be a peculiarity of the $d=6$ case. 
Given that this feature is irrelevant for the
main arguments of this work, I will not investigate it further.]
For $d=7$ and larger, I was unable to find any oscillons, 
confirming that $d_c=6$ for SDW potentials.

\begin{figure}[t]
\includegraphics[width=3in,height=3in]{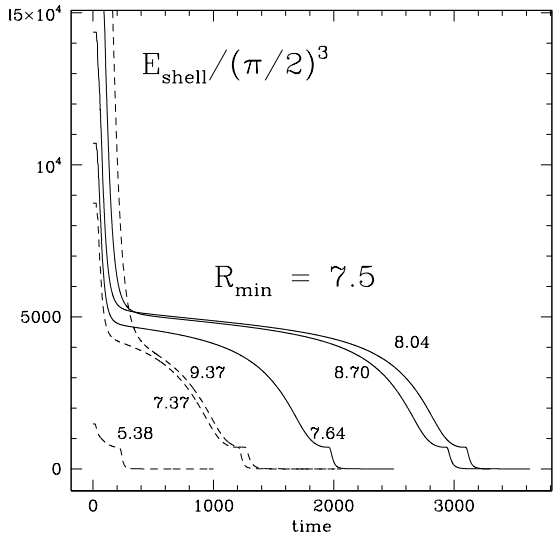}
\caption[oscillons for d=4]
{Time evolution of the energy within a shell of radius $R_{\rm shell} =10R$
in $d=6$. The labels specify the initial value of $R_{\rm eff}$. Continuous lines are for 
oscillons; dashed lines are for failed configurations.}
\label{d6_oscil}
\end{figure}

{}For an ``upside-down'' SDW, $g_4<0$. Writing the potential as 
$V(\phi) = \frac{1}{2}m^2\phi^2 - \frac{1}{4!}\phi^4$, and scaling
the field as $X = \phi/\phi_v$, with $\phi_v = \sqrt{6/\lambda}m$,
oscillons will exist as long as the amplitude of oscillations satisfies
$X^2 > 2^{d/2}\left(1+d/R^2\right)X_{\rm inf}^2$. $X_{\rm inf}$ is the
inflection point.

{}For asymmetric double-well potentials the situation changes.
From eq. (\ref{max-d}) it is easy to see that the 
upper critical dimension $d_c$
can vary as a function of the asymmetry:
the larger the absolute value of the
asymmetry, the higher $d_c$. As an
illustration, consider the potential
\begin{equation}
\label{ADWP}
V(\phi) = \frac{1}{2}\phi^2 - \frac{\alpha}{3}\phi^3 + \frac{1}{4}\phi^4~,
\end{equation}
\noindent
where $\alpha>0$ and the variables are all dimensionless. Oscillons with
lifetimes of order $10^4$m$^{-1}$ have been found for this model 
with $d=3$ \cite{copeland}.
The necessary condition eq. (\ref{cond-pot}) gives $\alpha^2 > 3$, which is the
same that guarantees an inflection point for $V(\phi)$. 
The condition for upper critical dimension reads, 
$d\leq {\rm Int}[\ln (3/\alpha^2)/\ln(2^{3/2}/3)]$. For example,
$\alpha^2 =9/2$, equivalent
to a SDW, gives $d_c=6$ as it should. $\alpha^2=5$ gives $d_c=8$.
Thus, asymmetries may relax (but not eliminate) the bound on the upper 
critical dimension for oscillons.

{}For SDW potentials, one can introduce dimensionless variables
$r' \equiv \sqrt{\lambda}\phi_v r$, $t' \equiv \sqrt{\lambda}\phi_v t$,
$X = \phi/\phi_v$ such that the energy scales as $E[\phi] = 
\lambda^{(2-d)/2}|\phi_v|^{4-d}E[X]$. From the numerical
results obtained, a rough (within a factor of 2)
estimate of their energies in $d$
dimensions is
$E[X]\sim \left(\frac{\pi}{2}\right)^{d/2}\frac{1}{2}d^{d-1}$. 
Of course, it's always
possible to obtain accurate results numerically, as shown in Figs. 1 and 2
(in units of $c_d$)
for $d=4$ and $d=6$, respectively: an oscillon in 5 dimensional
spacetime would have an energy of 
$E/c_5 \simeq 20 \lambda^{-1}$. 
The characteristic length-scale
of $d$-dimensional oscillons is determined by eq. (\ref{minrad SDW}).
Again, a rough estimate gives, 
$R_{\rm min}\sim \frac{d}{\sqrt{\lambda}\phi_v}$.
It is straightforward to
extend these arguments to arbitrary potentials.

The results of this work have established that if a potential
can support oscillons, they will have a well-defined
set of properties which are dimensionally-dependent: their energies
[the approximately flat plateaus of Figs. 1 and 2] and their average radii.
Also, the minimum radius for the initial configurations that lead to
oscillons is determined by the dimensionality of space, as seen for
the SDW potential in eq. (\ref{minrad SDW}). Thus, one can envision
that if such configurations were to be observed, and if the interactions
were known, their energies and sizes would uniquely determine the
dimensionality of space. In the example of Fig. 2
above ($d=6$), with a vacuum scale of $1$ TeV, a typical oscillon
will have a radius $R_{\rm eff} \sim 10^{-17}$ cm, while 
$R_{\rm KK} \sim 5\times 10^{-7}$. In this case, a flat space approximation 
such as the one used here would be quite acceptable, and the 
observed masses would receive only slight corrections from the extra dimensions.
A next step would be to examine if these configurations
exist for models with several interacting fields, including those
carrying Abelian and non-Abelian quantum numbers. It may be 
possible to find long-lived $d$-dimensional
$Q$-balls in $\varphi^4$ models. If $d=3$, this includes the
Standard Model and its supersymmetric extensions. Also,
an estimate of the oscillon lifetime and its dependence on
spatial dimensionality is still lacking. 
(Note how $d=6$ oscillons live four to
five time longer than those in $d=4$.)

Relaxing the constraint of having static, 
spatially-localized solutions to the equations of motion opens many avenues for 
further investigation: as was shown in this work,
long-lived {\it time-dependent} localized configurations
are supported by a wide class of models. They may not only be observed in $d=3$ but
also offer a new window into the extra dimensions, in case they exist.

I would like to thank Robert R. Caldwell
and Hans Reinhard-M\"uller for useful suggestions, and 
the National Science Foundation for partial financial support under grant 
PHYS-0099543.

 \end{document}